# Semi-nonparametric Latent Class Choice Model with a Flexible Class Membership Component: A Mixture Model Approach

## A Preprint

**Georges Sfeir**
American University of Beirut, Riad el Solh
1107 2020, Beirut, Lebanon
*E-mail: gms12@mail.aub.edu*

**Maya Abou-Zeid**
American University of Beirut, Riad el Solh
1107 2020, Beirut, Lebanon
*E-mail: ma202@aub.edu.lb*

**Filipe Rodrigues**
Technical University of Denmark (DTU),
Bygning 116B, 2800 Kgs. Lyngby, Denmark
*E-mail: rodr@dtu.dk*

**Francisco Camara Pereira**
Technical University of Denmark (DTU),
Bygning 116B, 2800 Kgs. Lyngby, Denmark
*E-mail: camara@dtu.dk*

**Isam Kaysi**
American University of Beirut, Riad el Solh
1107 2020, Beirut, Lebanon
*E-mail: isam@aub.edu.lb*

## ABSTRACT

This study presents a semi-nonparametric Latent Class Choice Model (LCCM) with a flexible class membership component. The proposed model formulates the latent classes using mixture models as an alternative approach to the traditional random utility specification with the aim of comparing the two approaches on various measures including prediction accuracy and representation of heterogeneity in the choice process. Mixture models are parametric model-based clustering techniques that have been widely used in areas such as machine learning, data mining and pattern recognition for clustering and classification problems. An Expectation-Maximization (EM) algorithm is derived for the estimation of the proposed model. Using two different case studies on travel mode choice behavior, the proposed model is compared to traditional discrete choice models on the basis of parameter estimates' signs, values of time, statistical goodness-of-fit measures, and cross-validation tests. Results show that mixture models improve the overall performance of latent class choice models by providing better out-of-sample predication accuracy in addition to better representations of heterogeneity without weakening the behavioral and economic interpretability of the choice models.

# 1 INTRODUCTION

Predicting people's choices remains a complex and challenging task. Modeling and understanding human decision-making are crucial for estimating the impact of new policies or services, especially within the transportation field. All around the globe, there are concerns regarding the consequences of high levels of traffic congestion, parking demand, vehicular emissions, etc. To develop sustainable travel policies that lessen the negative impacts of the transport system, it is crucial to understand behavioral patterns of commuters and forecast their travel mode choices with respect to changes in attributes of the transportation system (Bhat and Lawton, 2000).

Moreover, the digital revolution is reshaping every aspect of our life including the way we travel. New modes of transport, from car- and bike-sharing to Mobility on Demand (MOD) and Demand-Responsive Transit (DRT) services, are emerging as alternatives to classic public transportation systems with fixed routes and timetables. In addition, given the rapid growth rate at which the motor industry and its relevant technologies are evolving, autonomous and connected vehicles are expected to become commercially available in the near future. Predicting the impacts of such new modes on travel demand and mobility patterns is of utmost concern to researchers, transportation planners, policymakers, and operators alike.

Modeling and forecasting the demand for travel modes are usually done using discrete choice models (DCM), such as the multinomial logit model (MNL) and its variants, which are rooted in the traditional microeconomic theory of consumer behavior. These models assume that each decision-maker associates a utility to each available alternative and selects the alternative with the highest utility. The utility of an alternative is usually specified as a linear-in-the-parameters function of the alternative attributes and socio-economic characteristics of the decision-maker, in addition to a random term that represents the effect of unobserved variables.

This research develops a new semi-nonparametric discrete choice model that formulates the class membership component of Latent Class Choice Models (LCCM) as a mixture model, a method commonly used as an unsupervised technique in the machine learning community, to better estimate and predict the decision-making process of people when faced with different choice alternatives. LCCM is a nonparametric random-utility model used to identify behavioral heterogeneity by allocating individuals probabilistically to a set of homogenous latent classes. On the other hand, mixture models are parametric model-based clustering techniques that have been widely used in the last decades in several areas such as machine learning, data mining, pattern recognition, image analysis and several other clustering and classification problems (Viroli and McLachlan, 2019). Using two different mode choice applications, the proposed model is compared to traditional discrete choice models such as, LCCM, MNL, and continuous mixed logit models, on the basis of parameter estimates' signs, values of time, statistical goodness-of-fit measures, and cross-validation tests.

The remainder of this paper is organized as follows. First, we review the literature on discrete choice models and machine learning. Second, we present the mathematical formulation of the proposed hybrid model. Next, we present and compare the estimation results of two different choice case studies. Finally, we summarize our findings and discuss future extensions of this work.

## 2  LITERATURE

We start by reviewing discrete choice models and McFadden's formulation before discussing the concept of taste heterogeneity. Next, we review studies which have used machine learning techniques in travel mode choice modeling and the ones that have tried to combine machine learning with econometric models.

### 2.1  Discrete Choice Model

Discrete choice models derived from random utility maximization theory have been widely used to model choices made by decision-makers among a finite set of discrete alternatives. These models are used in different fields such as transportation, economics, finance, marketing, medicine, etc. Early forms of random utility maximization models were developed during the 1960s. However, it was McFadden's contribution to discrete choice analysis during the 1970s, the conditional logit model (McFadden, 1974), that received more attention from econometricians and researchers (Brathwaite et al., 2017; Manski, 2001). This is mainly due to the fact that he linked his MNL formulation to the classical consumer demand theory (McFadden, 2001).

According to McFadden's formulation, any econometric behavioral model should fulfill four main properties (Manski, 2001). First, the model should be consistent with utility theory, meaning that a decision-maker *n* facing a finite set of alternatives would select the alternative that maximizes his/her utility. Second, researchers must be able to forecast decision-makers' choices under different/new conditions and/or in different populations. This is achieved by defining the utility $U_{nj}$, that a decision-maker *n* might gain from choosing alternative *j*, as a function of some observed attributes of alternative *j* $(X_{nj})$ and characteristics of decision-maker *n* $(S_n)$. Third, the econometric analysis should account for the fact that the researcher will generally not be able to observe all aspects of the utility. Typically, some attributes of the alternatives and characteristics of the decision-makers will be missing from the data in-hand. Therefore, utility $U_{nj}$ is decomposed into two parts, a systematic utility $V_{nj}$ and a random disturbance term $\varepsilon_{nj}$. The systematic utility $V_{nj}$ also known as representative utility, is the product of the observed components ($X_{nj}$ and $S_n$) and a vector of unknown parameters $\beta$ that need to be estimated statistically using the available data. The disturbance $\varepsilon_{nj}$, a random term with a specific density hypothesized by the modeler, accounts for the contribution of the unobserved factors. Once the distribution of $\varepsilon_{nj}$ is fully specified, the researcher can estimate the probabilities of the decision-makers' choices. Finally, the econometric model should be computationally practical (Manski, 2001).

Researchers have trusted this MNL formulation due to its connection to consumer theory, closed-form choice probabilities, and simple interpretability. However, logit models suffer from strict statistical assumptions, such as the independence of irrelevant alternatives (IIA) which leads to proportional substitution patterns across alternatives (Train, 2003). While the IIA assumption captures people's behavior accurately in some situations, it might generate biased demand estimates in many other applications. In addition, the logit model can only represent taste variations (differences in choice behavior among individuals) when heterogeneity in the choice process varies systematically and not randomly and can only deal with panel data (i.e., data collected from the same individuals over time) when unobserved factors are uncorrelated over time and individuals.

During the last decades, different advanced discrete choice models have been developed to relax the behavioral limitations of the MNL while concurrently satisfying the above four properties. However, the field of discrete choice modeling still struggles with the question of how to better represent heterogeneity in the choice process (Vij and Krueger, 2017). Heterogeneity is known as taste variation across decision-makers and is usually captured through systematic or random specifications. When tastes vary systematically with observable variables, heterogeneity in the choice process is represented through interactions between socioeconomic characteristics related to the decision-makers and attributes of the alternatives. However, systematic specifications can lead to false conclusions, unreliable parameter estimates, and incorrect forecasts in case tastes vary randomly across decision-makers or are related to unobserved variables (Gopinath, 1995; Vij et al., 2013). Random taste heterogeneity is typically captured through mixed logit models which can approximate any random utility model (McFadden and Train, 2000). Mixed logit probabilities are defined as a weighted average of standard logit probabilities evaluated over a mixing distribution of parameters. This specification allows for different tastes/coefficients within the population. Most mixing distributions fall typically under two categories: parametric (also known as continuous mixed logit) and nonparametric distributions. Parametric distributions have predefined forms (e.g., normal, lognormal, etc.) with fixed parameters and usually provide great fit to the data. However, the choice of a proper distribution can be complicated and computationally expensive. Researchers have to make a prior assumption about the proper distribution or estimate different models with different distributions and then choose the best model based on statistical goodness-of-fit measures and behavioral interpretation of the parameter estimates (Vij and Krueger, 2017). Moreover, most of the parametric models estimated in the literature are limited to univariate distributions although some studies have tried to use mixture of continuous distributions as a random taste parameter distribution (e.g. Fosgerau and Hess, 2009; Keane and Wasi, 2013).

To overcome these constraints, researchers have relied on nonparametric distributions which do not have predefined shapes and do not require the researcher to make certain assumptions regarding the distributions of parameters across decision-makers. LCCM remains the most known and used nonparametric distribution. It is a random-utility model that extends the multinomial logit model by using the concept of latent class formulation and allows capturing heterogeneity in the choice process by allocating people probabilistically to a set of $K$ homogeneous classes that differ behaviorally from each other. It is usually used when the modeler postulates that the unobserved heterogeneity can be represented by discrete latent classes such as segments of the population with varying tastes, different decision protocols adopted by individuals, and choice sets considered which may vary from one individual to another (Gopinath, 1995). Several studies have tried to compare both continuous mixed logit and LCCM from a theoretical and empirical perspective (Andrews et al., 2002; Greene and Hensher, 2003; Han, 2019; Hess et al., 2009, to name a few). To sum up, LCCM has some advantages over the parametric/continuous mixed logit. First, LCCM makes fewer statistical assumptions regarding the parameters' distribution form. Second, unobserved heterogeneity in continuous mixed logit models suffers from a lack of interpretability since it is not usually explained by explanatory variables, although it is possible (Greene et al., 2006), while discrete latent classes are easily explained and interpreted since the class membership model of LCCM is usually a function of socio-economic characteristics of the decision-makers. Third, correlation between taste coefficients and elasticities are two major differences between the two approaches. In continuous mixed logit models, correlation can be accounted for by specifying a joint distribution for taste coefficients; however, most applications rely on independently distributed random taste coefficients. As for LCCM, correlation between taste coefficients is implicit in the model and it is a function of the class membership probabilities, which are a function of the socio-economic variables, and the class-specific taste coefficients. The same rationale applies to the elasticities which are not easily determined in continuous mixed logit models but are directly inferred from the class membership probabilities of LCCM. One major shortcoming of LCCM is that the discrete latent representation may oversimplify the unobserved heterogeneity, especially when a small number of classes is estimated, since the latent class is defined as a linear-in-the-parameters function of the socio-economic characteristics of the decision-makers.

In order to loosen some of the restrictions of parametric and nonparametric models, several studies have developed semi-nonparametric approaches with the most common one being finite mixture of distributions. For example, Bujosa et al. (2010) developed a hybrid model, Latent Class-Random Parameter Logit model (LC-RPL), to combine the concepts of latent class and random taste coefficients. The model outperformed the traditional LCCM and continuous mixed logit models in terms of goodness-of-fit and in-sample predictions. However, the application was limited to two latent classes and a univariate normal distribution for one taste coefficient. A similar approach was implemented by Greene and Hensher (2013). The hybrid model had better goodness-of-fit measures than traditional LCCM and continuous mixed logit model but was also limited to two latent classes and one univariate triangular distribution for a taste coefficient. Fosgerau and Hess (2009) compared two semi-nonparametric approaches against four continuous mixed logit models with different continuous distribution functions (normal, lognormal, triangular, and SB). The first approach uses a Mixture of Distributions (MOD) to define the distributions of random taste parameters while the second one uses the Normal distribution as a base for the random parameters and extends it by adding a series approximation of Legendre polynomials. The MOD approach had a slight advantage over the second approach and the traditional mixed logit models. However, it had computational problems and it was not possible to estimate more than a mixture of two normal distributions. Krueger et al. (2018) presented a Dirichlet process mixture multinomial logit (DPM-MNL) model where Dirichlet process is used as a flexible nonparametric mixing distribution for the parameters' coefficients. Such approach does not require the analyst to specify the number of mixtures a priori. However, it generates unstructured representations of heterogeneity which affects the interpretability of the model. Train (2008) developed an Expectation-Maximization (EM) algorithm for the estimation of mixture of distributions in mixed logit models. However, the application was also limited to mixture of two independent distributions for each randomly distributed coefficient. Moreover, Train (2016) introduced a new logit-mixed logit model where he relied on logit specifications to define the mixing distribution of random parameters. The framework proved its capability to approximate the shape of any mixing distribution but placed additional burden on the analyst to specify the utility of the random parameters and the variables that represent the shape of their distributions.

Most of the aforementioned studies have used mixtures of distributions to represent the random distribution of taste coefficients. Moreover, the majority of research on random heterogeneity has focused on improving the flexibility of utilities and parameters' coefficients. Instead, in this paper, we present an alternative semi-nonparametric approach which consists of using a mixture of distributions to formulate the latent classes (rather than the choice model parameters) and improve their flexibility. In other words, instead of defining more complex distribution functions, we use a mixture of distributions to cluster decision-makers. In the machine learning community, this is known as mixture models and it is widely used as an unsupervised technique to cluster data into homogeneous groups/clusters. We aim to compare this approach to the traditional LCCM, in addition to MNL and continuous mixed logit models, in terms of its ability to estimate different classes and improve prediction accuracy while keeping model interpretability and being useful for policy testing and inferring economic indicators.

## 2.2 Machine Learning

Since mixture models are used widely in the machine learning community, for the sake of completeness, we also review studies that have applied machine learning techniques to mode choice modeling problems. In recent years, supervised machine learning techniques have been used in mode choice modeling as alternative methods to traditional discrete choice models. Most of the contrast studies have shown that ML outperforms DCM in terms of prediction accuracy (Andrade et al., 2006; Cantarella and de Luca, 2005; Lee et al., 2018; Nijkamp et al., 1996; Sekhar et al., 2016; Tang et al., 2015; Xian-Yu, 2011; Xie et al., 2003). However, machine learning techniques tend to suffer from lack of interpretability and it is believed that this is the main reason that kept econometricians and transportation researchers from trusting such methods. Nevertheless, some efforts have been made to combine the two approaches. Gazder and Ratrout (2016) developed a sequential logit-Artificial Neural Network (ANN) framework for mode choice modeling and investigated its performance in different existing and hypothetical situations. It was found that single logit models have slightly better accuracies in predicting binary choice problems while the integrated logit-ANN approach performs better in multinomial choice situations. Sifringer et al. (2018) also proposed a hybrid sequential approach to enhance the predictive power of a logit model. The approach consists of adding an extra term, estimated by a Dense Neural Network (DNN), in the utilities of the original logit model. This extra term is estimated separately by using all disregarded variables in the logit model as input to the DNN model. This framework increased the final log-likelihood, the maximum joint probability of the observed dependent variables given the estimated parameters, by more than 15% while keeping the original DCM parameters statistically significant. Wong et al. (2018) explored the use of restricted Boltzmann machine (RBM), a generative non-parametric machine learning approach, to estimate latent variables without relying on measurement/attitudinal indicators as in Integrated Choice and Latent Variable (ICLV) models. Yafei (2019) developed a nonlinear-LCCM by using a neural network to specify the class membership model. The proposed model with 8 latent classes outperformed the best LCCM with 6 latent classes in terms of prediction accuracy. However, the nonlinear-LCCM is less transparent and loses interpretability at the latent class level due to the black-box nature of neural networks.

While most of the previous studies have focused on applying supervised machine learning (classification) to mode choice modeling or combining the two approaches in sequential approaches, this research aims at embedding unsupervised machine learning (clustering) in an econometric framework that satisfies McFadden's vision of a proper choice model. Clustering methods are used to discover heterogeneous subgroups or latent classes within a population by allocating similar observations (e.g. individuals with similar socio-demographic characteristics) to the same class/cluster. Different clustering techniques can be used including heuristics, hierarchical, $k$-means, and model-based clustering. We opt to use model-based clustering which is based on parametric mixture models. In such methods, each observation is assumed to be generated from a finite mixture of distributions where each distribution represents a latent class/cluster (McLachlan et al., 2019). The rationale for using model-based clustering in this study, as opposed to other techniques, is threefold. First, a probabilistic method is needed to estimate the proposed latent class – choice model framework simultaneously as opposed to a two-stage sequential approach. The simultaneous estimation usually provides more efficient estimates than the sequential estimation. Second, mixture models allow more flexibility than the utility specification of latent classes which is usually defined to be linear in parameters. Third, such techniques provide a framework for evaluating the clusters, meaning that interpretability can be maintained to a large extent (Biernacki et al., 2000).

## 3 MODEL FRAMEWORK AND FORMULATION

We develop a hybrid model that consists of using Gaussian Mixture Models (GMMs), a model-based clustering approach, as a first-stage clustering tool to divide the population into homogenous groups/classes while utilizing discrete choice models to develop class-specific choice models.

Gaussian Mixture Models are widely used in machine learning, statistical analysis, pattern recognition, and data mining and can be easily formulated to define discrete latent variables (Bishop, 2006). GMM is a combination of $K$ Gaussian densities where each density is a component (latent class) of the mixture and has its own mean vector and covariance structure. These models are more flexible than other clustering techniques (e.g. $k$-means or hierarchical clustering) since the covariance matrix of GMM can account for correlation between explanatory variables and clusters using different structures (McNicholas and Murphy, 2010). Particularly, the covariance matrices of GMM can have one of the following four structures: full covariance structure wherein each latent class has its own general covariance matrix, a tied covariance structure wherein all latent classes share the same general covariance matrix, a diagonal covariance structure wherein each latent class has its own diagonal covariance matrix, or a spherical structure wherein each latent class has one single variance. We believe this flexible approach would help capture underlying behavioral heterogeneity and complex behavioral patterns within the population. However, GMM can only deal with continuous variables. Therefore, we rely on a joint Gaussian-Bernoulli Mixture Model to assign decision-makers probabilistically to different latent classes using both continuous and discrete socio-economic characteristics while we make use of random utility models (e.g. logit models) for class-specific choice models. The full model is called Gaussian-Bernoulli Mixture Latent Class Choice Models (GBM-LCCM). This is similar to the well-known LCCM that allows capturing heterogeneity in the choice process by allocating people to a set of $K$ homogeneous classes.

The next section presents the LCCM formulation while the subsequent section develops the formulation and estimation technique of the proposed Gaussian-Bernoulli Mixture Latent Class Choice Model (GBM-LCCM).

### 3.1 Latent Class Choice Model

LCCM consists of two sub-models, a class membership model and a class-specific choice model (Figure 1). The class membership model formulates the probability of a decision-maker belonging to a specific class, typically as a function of his/her characteristics. The utility of decision-maker $n$ belonging to class $k$ is specified as follows:

$$U_{nk} = S'_n \gamma_k + v_{nk} \tag{1}$$

Where $S_n$ is a vector of characteristics of decision-maker $n$ including a constant, $\gamma_k$ is a vector of corresponding unknown parameters that need to be estimated statistically using the available data, and $v_{nk}$ is a random disturbance term that is assumed to be independently and identically distributed (iid) Extreme Value Type I over decision-makers and classes.

The probability of decision-maker $n$ belonging to class $k$ is then expressed as follows:

$$P(q_{nk}|S_n, \gamma_k) = \frac{e^{S_n' \gamma_k}}{\sum_{k'=1}^{K} e^{S_n' \gamma_{k'}}} \tag{2}$$

$$q_{nk} = \begin{cases} 1 \; if \; decision-maker \; n \; belongs \; to \; class \; k \\ 0 \; otherwise \end{cases} \tag{3}$$

Conditioned on the class membership of the decision-maker, the class-specific choice model estimates the probability of choosing a specific alternative as a function of the observed exogenous attributes of the alternatives. The utility of individual $n$ choosing alternative $j$ during time period $t$, conditional on him/her belonging to class $k$, is specified in the following manner:

$$U_{njt|k} = X_{njt}' \beta_k + \varepsilon_{njt|k} \tag{4}$$

Where $X_{njt}$ is a vector of observed attributes of alternative $j$ during time period $t$ including a constant, $\beta_k$ is a vector of corresponding unknown parameters that need to be estimated statistically using the available data, and $\varepsilon_{njt|k}$ is a random disturbance term that is independently and identically distributed (iid) Extreme Value Type I over decision-makers, alternatives, and classes.

Conditional on class $k$, the probability of decision-maker $n$ choosing alternative $j$ during time period $t$ is expressed as follows:

$$P(y_{njt}|X_{njt}, q_{nk}, \beta_k) = \frac{e^{V_{njt|k}}}{\sum_{j'=1}^{J} e^{V_{nj't|k}}} \tag{5}$$

Where $J$ is the number of alternatives.

Let $y_n$ be a ($J \; x \; T_n$) matrix of all choices of individual $n$ during all time periods $T_n$ and consisting of choice indicators $y_{njt}$ defined below. Let $X_n$ be a matrix consisting of $J \; x \; T_n$ vectors of $X_{njt}$. Conditional on class $k$, the probability of observing $y_n$ is expressed as follows:

$$P(y_n|X_n, q_{nk}, \beta_k) = \prod_{t=1}^{T_n} \prod_{j=1}^{J} \left( P(y_{njt}|X_{njt}, q_{nk}, \beta_k) \right)^{y_{njt}} \tag{6}$$

$$y_{njt} = \begin{cases} 1 \; if \; decision-maker \; n \; chooses \; alternative \; j \; during \; time \; period \; t \\ \quad 0 \; otherwise \end{cases} \tag{7}$$

The unconditional probability (or likelihood) of the observed choice of individual *n* can be obtained by mixing the conditional choice probability over the probability of belonging to each class *k*:

$$P(y_n) = \sum_{k=1}^{K} P(q_{nk}|S_n, \gamma_k) P(y_n|X_n, q_{nk}, \beta_k) \tag{8}$$

Finally, the likelihood over all decision-makers *N* is formulated as follows, assuming the availability of a sample of independent decision-makers:

$$P(y) = \prod_{n=1}^{N} \sum_{k=1}^{K} P(q_{nk}|S_n, \gamma_k) P(y_n|X_n, q_{nk}, \beta_k) \tag{9}$$

## 3.2 Gaussian-Bernoulli Mixture Latent Class Choice Model

We propose to replace the class membership model, $P(q_{nk}|S_n, \gamma_k)$, by a Gaussian-Bernoulli Mixture Model (GBM), a probabilistic machine learning approach used for clustering (Figures 2 and 3) where a Gaussian Mixture Model (GMM) is used for continuous variables and a Bernoulli Mixture Model (BMM) for discrete/binary variables. We split the vector of characteristics of decision-maker *n* ($S_n$) into two sub-vectors, $S_{cn}$ and $S_{dn}$. $S_{cn}$ accounts for the continuous characteristics of decision-maker *n* with dimension $D_c$ equal to the number of elements in $S_{cn}$ while $S_{dn}$ accounts for the discrete/binary characteristics of decision-maker *n* with dimension $D_d$ equal to the number of elements in $S_{dn}$.

GMM is a combination of *K* Gaussian densities where each density, $\mathcal{N}(S_{cn}|\mu_{ck}, \Sigma_{ck})$, is a component of the mixture and has its own mean $\mu_{ck}$ (with dimension equal to the number of elements in $S_{cn}$), covariance $\sum_{ck}$, and mixing coefficient $\pi_k$ (the overall probability that an observation comes from component *k*) (Bishop, 2006). BMM is a combination of *K* mixture components where each component *k* is a product of $D_d$ independent Bernoulli probability functions and has its own mean vector $\mu_{dk}$.

Replacing the class membership probability by a GBM is not a straightforward task. The probability of decision-maker *n* belonging to class *k*, $P(q_{nk}|S_n)$, is the posterior probability of the GBM and cannot be part of the likelihood function that needs to be maximized. Instead, we estimate the probability of observing decision-maker *n* with characteristics $S_n = \{S_{cn}, S_{dn}\}$ given that he/she belongs to latent class *k* (Figure 2). This stems from the fact that Gaussian-Bernoulli Mixture Models are generative models that learn the joint probability of the features/characteristics ($S_n$) and the labels/classes ($q_{nk}$) then make use of Bayes' theorem to calculate the posterior probability $P(q_{nk}|S_n)$ (Bishop, 2006). We follow the same steps to estimate the proposed GBM-LCCM. First, we estimate the joint probability of the model then we calculate the posterior and marginal probabilities by using Bayes rules. The graphical representation of the proposed hybrid model is shown in Figure 3.

Assuming that the continuous and binary data of the Gaussian and Bernoulli distributions are independent, the joint probability $S_{cn}$, $S_{dn}$, $y_n$ and $q_{nk}$ can be specified as the product of the class probability (first term on the right hand side below), the densities of $S_{cn}$ and $S_{dn}$ conditional on the class (second and third terms) and the choice probability conditional on the class (fourth term), as follows:

$$P(S_{cn}, S_{dn}, y_n, q_{nk}) = P(q_{nk}|\pi_k)P(S_{cn}|q_{nk} = 1, \mu_{ck}, \Sigma_{ck})P(S_{dn}|q_{nk} = 1, \mu_{dk})$$
$$\times P(y_n|X_n, q_{nk}, \beta_k)$$

$$(10)$$

Where:

$$P(q_{nk}|\pi_k) = \pi_k \tag{11}$$

$$\sum_{k=1}^{K} \pi_k = 1 \tag{12}$$

$$P(S_{cn}|q_{nk} = 1, \mu_{ck}, \Sigma_{ck}) = \mathcal{N}(S_{cn}|\mu_{ck}, \Sigma_{ck})$$
$$= \frac{1}{\sqrt{(2\pi)^{D_c}|\Sigma_{ck}|}} exp\left(-\frac{1}{2}(S_{cn} - \mu_{ck})'\Sigma_{ck}^{-1}(S_{cn} - \mu_{ck})\right) \tag{13}$$

$$P(S_{dn}|q_{nk} = 1, \mu_{dk}) = \prod_{i=1}^{D_d} \mu_{dk_i}^{S_{dni}}(1 - \mu_{dk_i})^{(1-S_{dni})} \tag{14}$$

With $|\Sigma_{ck}|$ the determinant of the covariance matrix, $S_{dni}$ a binary characteristics of decision-maker $n$ and $\mu_{dki}$ its corresponding mean.

The joint probability of $S_{cn}$, $S_{dn}$ and $y_n$ can be then obtained by taking the marginal of expression (10) over all components $K$:

$$P(S_{cn}, S_{dn}, y_n) = \sum_{k=1}^{K} P(S_{cn}, S_{dn}, y_n, q_{nk}) \tag{15}$$

Finally, the likelihood function of the proposed hybrid model for all decision-makers $N$ is formulated as follows:

$$P(S_c, S_d, y) = \prod_{n=1}^{N} P(S_{cn}, S_{dn}, y_n)$$
$$= \prod_{n=1}^{N} \sum_{k=1}^{K} \pi_k \mathcal{N}(S_{cn}|\mu_{ck}, \Sigma_{ck}) \prod_{i=1}^{D_d} \mu_{dk_i}^{S_{dni}}(1 - \mu_{dk_i})^{(1-S_{dni})}$$
$$\times \prod_{t=1}^{T_n} \prod_{j=1}^{J} \left(\frac{e^{X'_{njt}\beta_k}}{\sum_{j'=1}^{J} e^{X'_{nj't}\beta_k}}\right)^{y_{njt}} \tag{16}$$

Usually, traditional discrete choice models are estimated using maximum likelihood estimation techniques which aim at maximizing the likelihood of the observed data given the model parameters. However, maximizing the log-likelihood of both LCMM and GBM-LCCM is a complex task due to the summation over $k$ that will appear inside the logarithm of equations 9 and 16. Setting the derivatives of the log-likelihood to zero will not lead to a closed-form solution (Bishop, 2006). In addition, standard maximum likelihood estimation of nonparametric models (e.g. LCCM) becomes more difficult and requires significantly more time as the number of parameters increases. Moreover, the process of inverting the hessian matrix becomes numerically challenging as empirical singularity issues might arise at some iterations (Train, 2008). To overcome this, we refer to the Expectation-Maximization (EM) algorithm (Dempster et al., 1977), a powerful method used for maximum likelihood estimation in models with latent variables.

### 3.2.1 EM Algorithm

The first step of the EM algorithm requires writing the joint likelihood function (equation 16) assuming that the clusters (latent classes, $q_{nk}$) are observed:

$$P(S_c, S_d, y, q) = \prod_{n=1}^{N} \prod_{k=1}^{K} \left[ \pi_k \mathcal{N}(S_{cn}|\mu_{ck}, \Sigma_{ck}) \prod_{i=1}^{D_d} \mu_{dk_i}^{S_{dni}} (1 - \mu_{dk_i})^{(1-S_{dni})} \right]^{q_{nk}}$$
$$\times \prod_{n=1}^{N} \prod_{k=1}^{K} \prod_{t=1}^{T_n} \prod_{j=1}^{J} \left[ \frac{e^{X'_{njt}\beta_k}}{\sum_{j'=1}^{J} e^{X'_{nj't}\beta_k}} \right]^{y_{njt}q_{nk}} \tag{17}$$

Taking the logarithm of the likelihood, the function breaks into two separate parts, one for each of the two sub-models (class membership model and class-specific choice model), as follows:

$$LL = \sum_{n=1}^{N} \sum_{k=1}^{K} q_{nk} log \left[ \pi_k \mathcal{N}(S_{cn}|\mu_{ck}, \Sigma_{ck}) \prod_{i=1}^{D_d} \mu_{dk_i}^{S_{dni}} (1 - \mu_{dk_i})^{(1-S_{dni})} \right]$$
$$+ \sum_{n=1}^{N} \sum_{k=1}^{K} \sum_{t=1}^{T_n} \sum_{j=1}^{J} y_{njt} q_{nk} log \left[ \frac{e^{X'_{njt}\beta_k}}{\sum_{j'=1}^{J} e^{X'_{nj't}\beta_k}} \right] \tag{18}$$

Now, the unknown parameters $\{\mu_{ck}, \Sigma_k, \mu_{dk}, \pi_k, \beta_k\}$ of each component $k$ can be found by setting the derivatives of the above log-likelihood with respect to each of the unknown parameters to zero if and only if $q_{nk}$ is known. To find the values of $q_{nk}$, we estimate the expectation of $q_{nk}$ (E-step) using Bayes' theorem.

$$P(q_{nk}|y_n, S_{cn}, S_{dn}, X_n, \mu_{ck}, \Sigma_{ck}, \mu_{dk}, \pi_k, \beta_k)$$
$$\propto P(q_{nk}|\pi_k) P(S_{cn}|q_{nk}, \mu_{ck}, \Sigma_{ck}) P(S_{dn}|q_{nk}, \mu_{dk}) P(y_n|X_n, q_{nk}, \beta_k)$$
$$\propto \pi_k \mathcal{N}(S_{cn}|\mu_{ck}, \Sigma_{ck}) \prod_{i=1}^{D_d} \mu_{dk_i}^{S_{dni}} (1 - \mu_{dk_i})^{(1-S_{dni})} \prod_{t=1}^{T_n} \prod_{j=1}^{J} \left[ \frac{e^{X'_{njt}\beta_k}}{\sum_{j'=1}^{J} e^{X'_{nj't}\beta_k}} \right]^{y_{njt}} \tag{19}$$

$E[q_{nk}] = \gamma_{q_{nk}}$

$$= \frac{\pi_k \mathcal{N}(S_{cn}|\mu_{ck}, \Sigma_{ck}) \prod_{i=1}^{D_d} \mu_{dk_i}^{S_{dni}} (1-\mu_{dk_i})^{(1-S_{dni})} \prod_{t=1}^{T_n} \prod_{j=1}^{J} \left[\frac{e^{x'_{njt}\beta_k}}{\sum_{j'=1}^{J} e^{x'_{nj't}\beta_k}}\right]^{y_{njt}}}{\sum_{k'=1}^{K} \left[\pi_{k'} \mathcal{N}(S_{cn}|\mu_{ck'}, \Sigma_{ck'}) \prod_{i=1}^{D_d} \mu_{dk'_i}^{S_{dni}} (1-\mu_{dk'_i})^{(1-S_{dni})} \prod_{t=1}^{T_n} \prod_{j=1}^{J} \left[\frac{e^{x'_{njt}\beta_{k'}}}{\sum_{j'=1}^{J} e^{x'_{nj't}\beta_{k'}}}\right]^{y_{njt}}\right]} \quad (20)$$

It is to be noted that $\pi_k$ (equation 11) is considered as the prior probability of $q_{nk} = 1$ while $\gamma_{qnk}$ (equation 20) is the corresponding posterior probability.

Next, the likelihood should be maximized to find the unknown parameters. However, since equation 18 cannot be maximized directly due to the presence of latent variables $q_{nk}$, we consider instead the expected value of the log-likelihood, where the expectation is taken w.r.t. $q_{nk}$.

Making use of equations 18 and 20, gives:

$$\begin{aligned}
E[LL] = \sum_{n=1}^{N} \sum_{k=1}^{K} \gamma_{q_{nk}} &\left( log\pi_k + log\mathcal{N}(S_{cn}|\mu_{ck}, \Sigma_{ck}) \right. \\
&\left. + \sum_{i=1}^{D_d} [S_{dni} log\mu_{dki} + (1-S_{dni})log(1-\mu_{dki})] \right) \\
&+ \sum_{n=1}^{N} \sum_{k=1}^{K} \sum_{t=1}^{T_n} \sum_{j=1}^{J} y_{njt}\gamma_{q_{nk}} log\left[\frac{e^{x'_{njt}\beta_k}}{\sum_{j'=1}^{J} e^{x'_{nj't}\beta_k}}\right]
\end{aligned} \quad (21)$$

Setting the derivatives of the expected log-likelihood with respect to the unknown parameters to zero, we obtain the solutions of the unknown parameters as follows:

$$\mu_{ck} = \frac{1}{N_k} \sum_{n=1}^{N} \gamma_{q_{nk}} S_{cn} \quad (22)$$

$$\Sigma_{ck} = \frac{1}{N_k} \sum_{n=1}^{N} \gamma_{q_{nk}} (S_{cn} - \mu_{ck})(S_{cn} - \mu_{ck})' \quad (23)$$

$$\mu_{dk} = \frac{1}{N_k} \sum_{n=1}^{N} \gamma_{q_{nk}} S_{dn} \quad (24)$$

$$\pi_k = \frac{N_k}{N} \quad (25)$$

$$\beta_k = argmax_{\beta_k} \sum_{n=1}^{N} \sum_{t=1}^{T_n} \sum_{j=1}^{J} y_{njt} \gamma_{q_{nk}} log \left[ \frac{e^{X'_{njt}\beta_k}}{\sum_{j'=1}^{J} e^{X'_{nj't}\beta_k}} \right] \tag{26}$$

Where we have defined:

$$N_k = \sum_{n=1}^{N} \gamma_{q_{nk}} \tag{27}$$

Equations 22 to 25 are the closed-form solutions of the Gaussian mean matrix, Gaussian covariance matrix, Bernoulli mean matrix, and mixing coefficients, respectively. As for the parameters $\beta_k$ (equation 26), no closed-form solution can be obtained. Instead, we resort to the gradient-based numerical optimization method BFGS (Nocedal et al., 1999).

To sum up, the EM algorithm alternates between the E-step and M-step until convergence is reached. First, we initialize the unknown parameters. Second, we estimate the expected values of the latent variables using equation 20 (E-step). Next, we update the values of the unknown parameters using equations 22 to 26 (M-step). Finally, we evaluate the log-likelihood using the current values of the unknown parameters and check for convergence. If the convergence criterion is not met we return to the E-step.

### 3.2.2  Final Likelihood

After reaching convergence, we evaluate the marginal probability *P(y)* of observing a vector of choices *y* of all decision-makers *N* as follows:

$$P(y) = \prod_{n=1}^{N} \sum_{k=1}^{K} P(q_{nk}|S_{cn}, S_{dn}, \mu_{ck}, \Sigma_{ck}, \mu_{dk}, \pi_k) \prod_{t=1}^{T_n} \prod_{j=1}^{J} \left( P(y_{njt}|X_{njt}, q_{nk}, \beta_k) \right)^{y_{njt}} \tag{28}$$

Where $P(q_{nk}|S_{cn}, S_{dn}, \mu_{ck}, \Sigma_{ck}, \mu_{dk}, \pi_k)$ is the posterior probability of vector $S_n = \{S_{cn}, S_{dn}\}$ being generated by cluster *k*. The posterior probability can be formulated using Bayes' theorem:

$$\begin{aligned} P(q_{nk}&|S_{cn}, S_{dn}, \mu_{ck}, \Sigma_{ck}, \mu_{dk}, \pi_k) \\ &= \frac{P(q_{nk}|\pi_k)P(S_{cn}|q_{nk}, \mu_{ck}, \Sigma_{ck})P(S_{dn}|q_{nk}, \mu_{dk})}{\sum_{k'=1}^{K} P(q_{nk'}|\pi_{k'})P(S_{cn}|q_{nk'}, \mu_{ck'}, \Sigma_{ck'})P(S_{dn}|q_{nk'}, \mu_{dk'})} \end{aligned} \tag{29}$$

The above marginal probability (equation 28) is used for comparing the GBM-LCCM with the traditional LCCM (equation 9) and for calculating out-of-sample prediction accuracies.

Note that, in case only continuous socio-economic characteristics are used, the proposed model becomes Gaussian Mixture Latent Class Choice Model (GM-LCCM). The formulation would follow the same steps of sections 3.2 but without the mixture of Bernoulli distribution functions (equation 14). The same applies in case only discrete variables are used in the clustering stage.

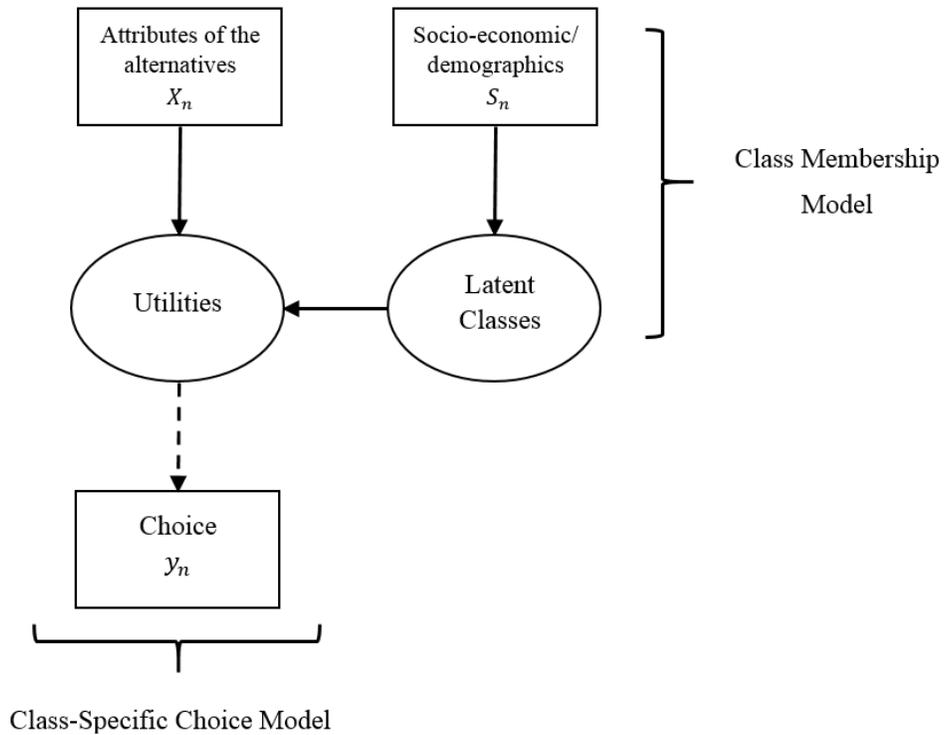

**Figure 1: Traditional Latent Class Choice Model (LCCM)**

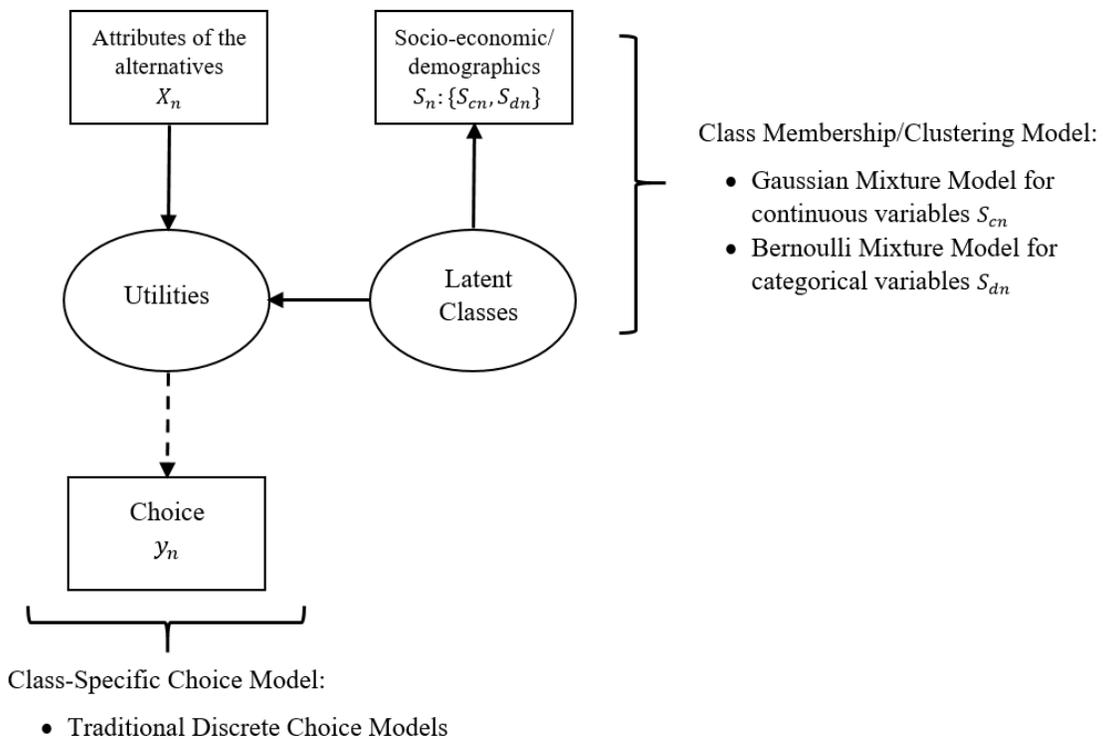

**Figure 2: Gaussian-Bernoulli Mixture Latent Class Choice Model (GBM-LCCM)**

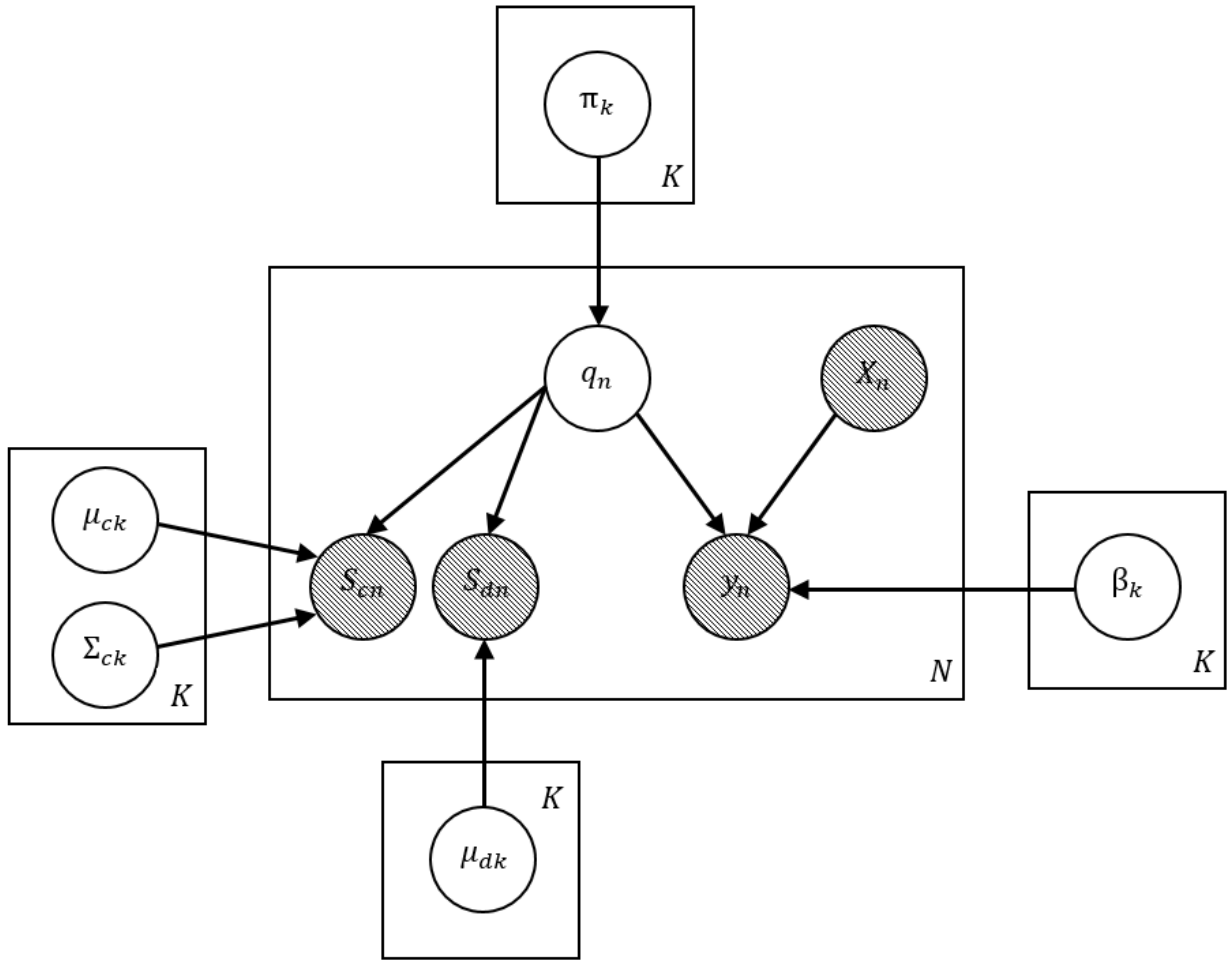

**Figure 3: Graphical Representation of the proposed Gaussian-Bernoulli Mixture Latent Class Choice Model (GBM-LCCM) for a set of N decision-makers and K clusters/classes**

## 4    APPLICATION

In this section we develop and present two applications of the proposed modeling approach (GBM-LCCM) using two different case studies on travel mode choice behavior. In addition, we estimate and present MNL, Mixed Logit, and LCCM models to benchmark the proposed GBM-LCCM against traditional discrete choice models. The proposed models (GBM-LCCM) are programmed in Python using some implementations from Scikit-Learn library (Pedregosa et al., 2011) and lccm package (El Zarwi, 2017) to implement the EM method presented in section 3.2. The traditional LCCMs are also estimated in Python using the lccm package (El Zarwi, 2017) which provides an EM algorithm for maximizing the likelihood function of traditional latent class choice models. The MNL and Mixed Logit models are estimated in PythonBiogeme (Bierlaire, 2016) using maximum and maximum-simulated likelihood methods, respectively.

The EM algorithm has proved to be a powerful approach for estimating models with latent variables or missing data (Bhat, 1997; Train, 2008). However, the algorithm is sensitive to starting values and might not guarantee convergence to the global maximum. Therefore, a good set of initial values is of great importance to assure proper convergence. Different approaches and heuristics have been used in the literature to overcome this limitation. The two most used approaches are random initialization and incremental initialization, where estimates of models with K-1 classes are used as starting values for models with K classes. In this study, we make use of both approaches (Table 1). In addition, the Gaussian-Bernoulli Mixture Models are initialized randomly and using k-means, a deterministic unsupervised machine learning approach. In total, each model is estimated 25 times and the variance of the log-likelihood is reported to check if the model is stable or not.

**Table 1: EM Initialization**

| GBM-LCCM | | | LCCM | | |
|---|---|---|---|---|---|
| GBM | Class Specific Choice Model | Number of Trials | Class Membership Model | Class Specific Choice Model | Number of Trials |
| Random | 0 | 5 | 0 | 0 | 1 |
| Random | Random | 5 | 0 | Random | 5 |
| K-means | 0 | 5 | Random | 0 | 5 |
| K-means | Random | 5 | Random | Random | 5 |
| Random/K-means | Estimates of K-1 model and 0/Random for the additional class | 5 | Estimates of K-1 model and 0/Random for the additional class | | 5 |

Next, we present and discuss the results of the two case studies.

### 4.1    Revealed Preferences (RP) Case Study

The first application is based on the dataset from Hillel et al. (2018) which is available online as a supplementary material to the paper. The dataset combines individual trip diaries of the London Travel Demand Survey (LTDS) from April 2012 to March 2015 with their corresponding modes alternatives extracted from a Google directions application programming interface (API) and corresponding estimates of car operating costs and public transport fares. The dataset consists of 81,086 trips, four modes (walking, cycling, public transport, and driving), and different trip purposes (e.g. Home-Based Work, Home-Based Education, etc.). In this application, we only consider Home-Based Work (HBW) trips and trips made by car and public transport in order to have a balanced sample. The first two years (7,814 trips) are used for estimation/training while the third year (3,883 trips) is used for testing/prediction.

We test and present three different specifications and, for the sake of brevity, we only present summary statistics of the estimated models.

### 4.1.1   First Specification

We assume that the latent classes of the GBM-LCCM are characterized by the available socio-economic variables age, gender, car ownership, and driving license, such that $age_n$ is a continuous variable representing the age of decision-maker $n$; $female_n$ is a binary variable that equals to 1 if decision-maker $n$ is female and 0 otherwise; $car\_own_{n1}$ is a binary variable that equals to 1 if the number of cars in the household of decision-maker $n$ is more than 0 but less than one per adult and 0 otherwise, $car\_own_{n2}$ equals to 1 if the number of cars in the household of decision-maker $n$ is one or more per adult and 0 otherwise; and $license_n$ is a binary variable that equals to 1 if decision-maker $n$ has a driving license and 0 otherwise. Since only one continuous variable (age) is used for clustering, two covariance structures, full and tied, are tested. Regarding the class-specific choice models, we only consider alternative-specific travel time and travel cost coefficients in addition to a constant in the utility of the car alternative.

Moreover, in order to compare the new approach with existing discrete choice models, we estimate four models: MNL model, mixed logit model wherein the travel time coefficients are normally distributed, mixed logit model wherein the travel time coefficients are lognormally distributed and traditional LCCM with the same specification as the GBM-LCCM.

Table 2 presents summary statistics of the new approach and the four traditional models. We enumerate the average joint log-likelihood of the GBM-LCCM models, the average marginal log-likelihood of all models, the corresponding Akaike Information Criterion (AIC) and Bayesian Information Criterion (BIC), the predictive log-likelihood (for the test sample), and the variance of the marginal log-likelihood of LCCM and GBM-LCCM models to evaluate the stability of the EM solutions since these models are run multiple times with different starting values.

We first look at the marginal log-likelihoods of all the estimated models. All mixture models (continuous and discrete) have better model fit than the MNL model, with LCCM (K = 3) having the best log-likelihood (-2,470.01). However, the log-likelihood variance of the LCCM with three latent classes is very high (190.64) meaning the model is highly unstable and should be neglected. Moreover, LCCM (K = 2), mixed logit with normal distributions, and mixed logit with lognormal distributions have positive public transport cost coefficients. These models are also ignored since travel cost coefficients should have a negative sign. For the LCCM approach, no more than three classes are estimated mainly due to identification problems (high standard deviations for the class-specific parameter estimates). Regarding the proposed GBM-LCCM, the full covariance model with 5 latent classes is also unstable (high variance) and should be ignored. After eliminating all unstable models and models with unexpected coefficients' signs, GBM-LCCM with a tied covariance structure and 4 latent classes can be selected as the best model since it has the best AIC, BIC, and predictive power. However, it is to be noted that models with tied and full covariance structures have similar performance, with slightly better log-likelihood for the model with a full covariance structure and slightly better predictive power for the model with a tied structure. This is mainly due to the additional parameters from the full covariance structure of the GMM.

Next, the latent classes of the tied-GBM-LCCM with four classes are described based on the mean matrix of the Gaussian-Bernoulli Mixture Model (Table 3). Note that the continuous variable age is standardized. Therefore, a negative value means the latent class is characterized by young individuals while a positive value means individuals are older than the average (which is 40 years).

*K1: Licensed drivers in their forties*

The first latent class is characterized by individuals with an age near the population average (40 years) since the mean of age is around 0. Individuals belonging to this class are mostly licensed drivers (91.9%) and living in households with low car ownership.

*K2: Young with low car ownership*

The second latent class has the youngest individuals ($\mu_{age} < 0$) from both genders who are almost equally likely to be licensed (48.8%) or unlicensed drivers (51.2%). In addition, individuals belonging to this class live in households with no cars (55.1%) or less than one car per adult (42.4%).

*K3: Licensed elderly*

This class includes the oldest individuals (highest $\mu_{age}$ across all classes) who are mostly males (72.1%), licensed drivers (94.5%), and belong to families with less than one car per adult (98.6%).

*K4: Licensed drivers with high car ownership*

The last latent class is characterized by old individuals from both genders. Moreover, individuals are licensed drivers (99%) who live in households with more than one car per adult (98.2%).

The above analysis is a strong indication that the proposed model provides a simple interpretability at the class membership level, although the random utility formulation of the latent classes is replaced by a full mixture model.

Finally, table 4 presents the class-specific parameter estimates of the tied-GBM-LCCM model with 4 classes. All cost and travel time parameters have the expected negative sign. Individuals from the first, third, and fourth classes are insensitive to travel cost of public transport.

*4.1.2   Second Specification*

In the second trial, we adopt the same class membership specification as in the first trial but a more complex class-specific choice utilities' specification. In particular, public transport travel time is included in the utilities as three separate attributes: access travel time (walking time between origin and first public transport stage, and final public transport stage and destination), bus/rail travel time (travel time spent on rail and bus services), and interchange travel time (walking and waiting time at the stop for interchanges on public transport route). However, all five models generated positive public transport cost coefficients. Therefore, a logarithmic specification of public transport cost was used in the class-specific choice utilities in order to resolve the issue of counter-intuitive sign of cost coefficients. Table 5 presents summary statistics of the new approach and the four traditional models. The logarithmic transformation of public transport cost did solve the issue of counter-intuitive signs for the GBM-LCCM with two and three latent classes but it had no impact on the remaining models. In addition, LCCM and GBM-LCCM models with positive cost coefficients showed very high convergence instability (very high log-likelihood variances). After eliminating all models with counter-intuitive coefficient signs and high log-likelihood variances, we can select the tied-GBM-LCCM with 3 latent classes as the best model since it has the best AIC, BIC, and predictive log-likelihood. Note that the results of this specification are consistent with that of section 4.1.1. Tied and full covariance structure models have similar performance with slight differences in goodness-of-fit and predictive measures due to the differences in the covariance structures of the GMM.

### 4.1.3    Third Specification

For the third and last attempt, we consider the same latent classes' formulation and class-specific choice utilities' specification as in the first trial (section 4.1.1). In addition, we include in the class-specific choice utilities four additional variables (start time, day of week, month, and traffic variability[1]). Traffic variability is added to the utilities as a continuous variable while the remaining three variables are binned and included as dummy variables. We use the same bins that are defined by Hillel et al. (2019). The start time of the trips is grouped into four categories: AM peak (06:30-09:29), inter-peak (09:30-16:29), peak (16:30-19:29), and night (19-30:06:29). The day of the week is divided into week days (Monday to Friday), Saturday, and Sunday. Finally, the trip month is grouped into winter season (December to February) and all other months (March to November). Table 6 presents summary statistics of all estimated models. LCCM ran into identification issues (class-specific choice parameter estimates with very large standard errors) while GBM-LCCM was able to determine only two latent classes. This might be an indication that the heterogeneity within the sample is rather systematic than random. Nevertheless, the Gaussian-Bernoulli mixture formulation of the latent classes showed a superior clustering ability by determining two homogeneous groups within the sample while the traditional random utility formulation of the LCCM had computational problems.

---

[1] Traffic variability is a measure of traffic variability for the driving route and it is calculated using travel times during optimistic, pessimistic, and typical traffic conditions, as predicted by the directions API. For more details, readers may refer to Hillel et al. (2018).

**Table 2: First specification**

| | K | Joint LL[a] | LL[b] | Variance[c] | AIC | BIC | Pred. LL | Notes |
|---|---|---|---|---|---|---|---|---|
| **MNL** | | | -4,017.47 | | 8,044.94 | 8,079.76 | -1,914.29 | |
| **Mixed Logit Normal** | | | -3,087.82 | | 6,189.64 | 6,238.38 | -1,896.77 | $\beta_{cost\_pt}$ = 0.0384 (p = 0.77) |
| **Mixed Logit LogNormal** | | | -3,090.49 | | 6,194.97 | 6,243.72 | -1,889.46 | $\beta_{cost\_pt}$ = 0.0956 (p = 0.41) |
| **LCCM** | 2 | | -2,643.15 | 0 | 5,318.30 | 5,429.72 | -1,234.47 | $\beta_{cost\_pt2}$ = 0.0618 (p = 0.14) |
| | 3 | | -2,470.01 | 190.64 | 4,994.02 | 5,182.04 | -1,182.95 | $\beta_{cost\_pt2}$ = 0.0748 (t = 0.56) |
| **GBM-LCCM Full Covariance** | 2 | -18,660.22 | -2,920.92 | 0 | 5,887.84 | 6,048.00 | -1,387.89 | |
| | 3 | -17,502.93 | -2,807.87 | 0 | 5,685.74 | 5,929.47 | -1,300.33 | |
| | 4 | **-17,390.22** | **-2,703.27** | 0 | 5,500.54 | 5,827.83 | -1,262.86 | |
| | 5 | -17,148.88 | -2,671.04 | 27.45 | 5,460.08 | 5,870.94 | -1,266.22 | $\beta_{cost\_pt1}$ = 0.0258 (p = 0.86) |
| **GBM-LCCM Tied Covariance** | 2 | -18,662.68 | -2,920.81 | 0 | 5,885.62 | 6,038.82 | -1,387.40 | |
| | 3 | -17,505.89 | -2,807.91 | 0 | 5,681.82 | 5,911.62 | -1,300.17 | |
| | 4 | -17,393.78 | -2,703.38 | **0** | **5,494.76** | **5,801.16** | **-1,260.08** | |

a: joint log-likelihood of the GBM-LCCM (equation 16)

b: marginal log-likelihood of the GBM-LCCM (equation 28) and the LCCM (equation 9)

c: variance of the marginal log-likelihood (LL)

**Table 3: Mean matrix of the class membership model (GBM) – Tied Covariance – K = 4**

| Variables | | Class 1 | Class 2 | Class 3 | Class 4 |
|---|---|---|---|---|---|
| **Age**[*] | Continuous | 0.034 | -0.285 | 0.447 | 0.332 |
| **Gender** | Female | 0.426 | 0.533 | 0.279 | 0.463 |
| | Male | 0.574 | 0.467 | 0.721 | 0.537 |
| **Driving License** | Yes | 0.919 | 0.488 | 0.945 | 0.990 |
| | No | 0.081 | 0.512 | 0.055 | 0.010 |
| **Household Car Ownership per Adult** | 0 | 0.056 | 0.551 | 0.014 | 0.018 |
| | ] 0 − 1 [ | 0.944 | 0.424 | 0.986 | 0 |
| | ≥ 1 | 0 | 0.025 | 0 | 0.982 |

[*]Age is standardized to have a mean of 0 and standard deviation of 1.

**Table 4: Parameter estimates of the class-specific choice models – Tied Covariance – K = 4**

| Variables | Class 1 | Class 2 | Class 3 | Class 4 |
|---|---|---|---|---|
| **ASC (Car)** | 2.354 (0.00) | -0.858 (0.00) | 2.513 (0.00) | 1.516 (0.00) |
| **Travel Time (PT)** | -0.178 (0.00) | -0.0751 (0.00) | -0.112 (0.00) | -0.0646 (0.00) |
| **Travel Time (Car)** | -0.316 (0.00) | -0.284 (0.00) | -0.115 (0.00) | -0.106 (0.00) |
| **Cost (PT)** | -0.102 (0.28) | -0.267 (0.01) | -0.106 (0.49) | -0.0206 (0.63) |
| **Cost (Car)** | -0.492 (0.00) | -0.181 (0.06) | -0.207 (0.00) | -0.153 (0.00) |

Values within parentheses are p-values.

**Table 5: Second specification**

| | K | Joint LL[a] | LL[b] | Variance[c] | AIC | BIC | Pred. LL | Notes |
|---|---|---|---|---|---|---|---|---|
| **MNL** | | | -4,003.98 | | 8,021.97 | 8,070.71 | -1,901.9 | $\beta_{Log\_cost\_pt}$= 0.0137 (p = 0.24) |
| **Mixed Logit Normal** | | | -3,020.68 | | 6,063.36 | 6,139.96 | -1,891.38 | $\beta_{Log\_cost\_pt}$= 0.0199 (p = 0.77) |
| **Mixed Logit LogNormal** | | | -3,035.05 | | 6,092.10 | 6,168.70 | -1,887.88 | $\beta_{Log\_cost\_pt}$= 0.0622 (t = 0.42) |
| **LCCM** | 2 | | -2,633.56 | 880.40 | 5,307.12 | 5,446.39 | -1,223.45 | $\beta_{Log\_cost\_pt2}$= 0.0134 (p = 0.55) |
| | 3 | | -2,458.11 | 750.66 | 4,982.22 | 5,212.02 | -1,192.74 | $\beta_{Log\_cost\_pt1}$= 0.188 (p = 0.00) |
| **GBM-LCCM Full Covariance** | 2 | -18,646.92 | -2,906.82 | 0 | 5,867.64 | 6,055.66 | -1,391.48 | |
| | 3 | **-17,485.77** | **-2,790.88** | 0 | 5,663.76 | 5,949.27 | -1,302.31 | |
| | 4 | -17,365.19 | -2,684.59 | 2,712.20 | 5,479.18 | 5,862.18 | -1,271.95 | $\beta_{Log\_cost\_pt1}$= 0.0779 (p = 0.33) |
| **GBM-LCCM Tied Covariance** | 2 | -18,650.17 | -2,907.00 | 0 | 5,866.00 | 6,047.06 | -1,390.03 | |
| | 3 | -17,489.31 | -2,791.21 | 0 | **5,660.42** | **5,932.00** | **-1,301.78** | |
| | 4 | -17,372.80 | -2,684.57 | 95.95 | 5,473.14 | 5,835.25 | -1,260.96 | $\beta_{Log\_cost\_pt2}$= 0.0234 (p = 0.82) |

a: joint log-likelihood of the GBMLCCM model (equation 16)

b: marginal log-likelihood of the GBMLCCM (equation 28) and the LCCM (equation 9)

c: variance of the marginal log-likelihood (LL)

**Table 6: Third specification**

| | K | Joint LL[a] | LL[b] | Variance[c] | AIC | BIC | Pred. LL | Notes |
|---|---|---|---|---|---|---|---|---|
| **MNL** | | | -3,778.40 | | 7,580.80 | 7,664.36 | -1,825.66 | |
| **Mixed Logit Normal** | | | -2,983.79 | | 5,995.58 | 6,093.07 | -1,864.30 | $\beta_{cost\_pt}$= 0.0376 (p = 0.77) |
| **Mixed Logit LogNormal** | | | -2,981.83 | | 5,991.66 | 6,089.15 | -1,850.34 | $\beta_{cost\_pt}$= 0.113 (p = 0.32) |
| **LCCM** | | | | | | | | Identification Issues |
| **GBM-LCCM Full Covariance** | 2 | -18,506.50 | -2,769.56 | 0.19 | 5,613.12 | 5,870.78 | -1,336.19 | |
| **GBM-LCCM Tied Covariance** | 2 | -18,508.80 | -2,769.17 | 0.61 | 5,610.34 | 5,861.03 | -1,335.17 | |

a: joint log-likelihood of the GBMLCCM model (equation 16)

b: marginal log-likelihood of the GBMLCCM (equation 28) and the LCCM (equation 9)

c: variance of the marginal log-likelihood (LL)

## 4.2 Stated Preferences (SP) Case Study

A second application of the proposed modeling approach is developed using a dataset from the American University of Beirut (AUB) in Lebanon, a major private university with about 8,094 students, 4,173 staff, and 2,168 faculty members (*AUB Fact Book 2016-2017*, 2016). The university is located in a dense urban area within Municipal Beirut and its surrounding neighborhood suffers from high levels of congestion and parking demand. To overcome these problems, AUB is considering two alternative sustainable transport modes for its population, shared-taxi and shuttle services. The shared-taxi would be a door-to-door service that provides on-demand transport between AUB gates and users' residences (and vice versa) while the shuttle service would be a non-stop first/last mile service between AUB gates and satellite parking hubs (and vice versa) where commuters could park their cars just a few kilometers away from AUB. In order to investigate the willingness of the AUB population to use the new transport services if they were implemented, a web-based stated preferences commuting survey was designed and sent to all AUB students, faculty members, and staff in April of 2017. The survey collected information about each respondent's daily travel to and from AUB, place of residence, and socio-economic characteristics. In addition, the stated preferences survey offered each respondent four hypothetical scenarios in which he/she had to state how many weekdays per week he/she is willing to use the two proposed services in addition to his/her current mode of commute. An example of the hypothetical scenarios is shown in Figure 4. A sub-sample of car users who come five days per week to AUB is used in this application. The sub-sample consists of 650 respondents and 2,600 choice observations. For more details about the dataset and the survey design, readers may refer to Sfeir et al. (2020).

In this application, we only consider mixed logit models (LCCM and GM-LCCM) in order to have a more in-depth comparison. We also only consider continuous variables for clustering in order to investigate the impact of the different covariance structures of GMM. We model the weekly frequency of commuting by three different modes (shared-taxi 'ST', shuttle 'SH', and current mode 'Car'). The choice data is multivariate count data with a fixed total count as the number of times that an individual commutes to AUB per week is fixed and exogenous. We use a full enumeration of all count combinations to model the choices (Ben-Akiva and Abou-Zeid, 2013; Sfeir et al., 2020). In such an approach, the universal choice set would consist of all possible combinations of weekly frequencies of using the three available modes. The weekly scheduling for those who travel 5 days per week to AUB involves a choice from a choice set consisting of 21 alternatives.

| Shared-Taxi | Shuttle | Your Current Commute to AUB |
|---|---|---|
| <u>Door-to-door travel time</u><br>**33 min**<br><u>Waiting time for late pick-up and/or early drop-off</u><br>**0 to 5 min**<br><u>Number of passengers sharing a ride</u><br>**4 to 6 (Minivan)** | <u>Access travel time</u><br>**12 min (by car)**<br><u>Frequency</u><br>**Every 5 min**<br><u>In-Shuttle travel time</u><br>**27 min** | <u>Total Travel Time</u><br>**30 min**<br><u>Mode of Travel</u><br>**Car** |
| <u>Mobile App/Wi-Fi/Live tracking</u><br>**Available** | <u>Wi-Fi/Live tracking</u><br>**Not Available** | |
| <u>One-way fare</u><br>**4,000 L.L.** | <u>One-way shuttle fare including parking cost</u><br>**1,000 L.L.**<br><u>One-way fuel cost (for access by car)</u><br>**700 L.L.** | <u>Parking cost</u><br>**5,000 L.L.**<br><u>One-way fuel cost</u><br>**1,800 L.L.** |

Based on this scenario, and considering your current pattern to AUB, how many weekdays per week will you use the proposed services? Remember that you indicated earlier that you come on 5 weekdays per week to AUB.

| Shared-Taxi | 0, 1, 2, 3, 4, 5 |
|---|---|
| Shuttle | 0, 1, 2, 3, 4, 5 |
| Your Current Commute to AUB | 0, 1, 2, 3, 4, 5 |

**Figure 4: Hypothetical scenario and choice question example from the survey**

The systematic utility of an alternative can then be specified as follows:

$$
\begin{aligned}
V_n\big(ST_i, SH_j, Car_k\big)_t = {} & C_{ST_i} + C_{SH_j} + C_{Car_k} \\
& + i \times \big(\beta_{Cost_{ST}} Cost_{n,ST,t} + \beta_{n,TT_{ST}} TT_{n,ST,t}\big) \\
& + j \times \big(\beta_{Cost_{SH}} Cost_{n,SH,t} + \beta_{n,TT_{SH}} TT_{n,SH,t} + \beta_{Head} Head_{n,SH,t}\big) \\
& + k \times \big(\beta_{Cost_{Car}} Cost_{n,Car,t} + \beta_{n,TT_{Car}} TT_{n,Car,t}\big)
\end{aligned}
\tag{30}
$$

Where $i$, $j$, and $k$ $(0, 1, 2, 3, 4, 5)$ are the number of weekly trips by shared-taxi, shuttle and car, respectively. The $Cs$ replace the traditional alternative-specific constants and are related to the frequencies of using the three modes per week. Eighteen constants (6 for each mode) have to be defined since the frequency of using each mode varies between 0 and 5 in a specific week. Four constants, $C_{ST0}$, $C_{SH0}$, $C_{Car0}$ and $C_{Car5}$, are set to zero for identification purposes. The specification assumes that the impact of time, headway, and cost variables on the utility is proportional to the number of times per week that a certain mode is used.

Table 7 shows the explanatory variables used in both models, LCCM and GM-LCCM. Several other variables such as income, household car ownership, and parking location were tested but they were insignificant. The coefficients of travel cost and travel time are specified as alternative (mode)-specific to account for variations in values of time (VOT) across users of different modes (Guevara, 2017).

Table 8 presents summary statistics of the LCCM and the GM-LCCM. For the LCCM, it was not possible to increase the number of latent classes beyond three. In doing so, LCCM generated very high standard errors for the class-specific parameter estimates. As for the GM-LCCM, there were no identification problems involved in increasing the number of latent classes up to five. However, GM-LCCM with higher number of latent classes (K > 2) resulted in positive travel cost and/or travel time coefficients, except for the spherical structure model with three latent classes, and thus these models are excluded from the comparison. Note that the full sub-sample, consisting of 650 respondents and 2,600 choice observations, is used for estimation. The predictive power of the models is compared using the 5-fold cross validation technique. The dataset is divided into 5 subsets and each model is trained 5 times. Each time, the models are trained on 4 different subsets and tested on the remaining one. Next, the log-likelihood of each of the test sets is calculated and the average value is reported. For the case of two latent classes (K = 2), results show that the tied structure model has similar marginal log-likelihood as the LCCM but a better prediction log-likelihood. This suggests that the GM-LCCM performs better in terms of prediction accuracy although both models have similar goodness-of-fit measure (LL). The three other covariance structures (full, diagonal, and spherical) have also a better prediction accuracy than the LCCM.

Tables 9 and 10 present estimates of the sub-models of LCCM and tied-GM-LCCM with two classes in addition to the VOT estimates (values between parentheses are p-values). The covariance estimates are not shown for conciseness. Results show that the estimates of the class-specific choice models of the two approaches are almost the same. All travel cost and travel time parameters have the expected negative sign. Members of the first class seem to be more sensitive to travel time. Results of the class membership model of the LCCM reveal that members of the first class are more likely young people and staff with low grades who live in households with fewer cars, and don't share rides to AUB. The signs of the means from the class membership model of the GM-LCCM lead to the same conclusion. Members of the second class have similar VOT for car and shuttle, which is also a trip by car where a user parks his/her car in a parking garage and uses the shuttle as a first/last mile service to/from AUB, while members of the first class have higher VOT for car. In terms of log-likelihood, both models have the same fitted value. We believe that the improvement in prediction accuracy (Table 8) is due to the changes in the class membership model since the parameter estimates of both class-specific choice models are almost the same (Tables 9 and 10).

Moreover, results of GM-LCCM with three latent classes and a spherical covariance structure are presented in Table 11. Individuals from the third class appear to be insensitive towards travel cost of car and travel time of shuttle, hence the high and low VOTs of car and shuttle, respectively. Going back to Table 8, it is clear that the GM-LCCM with three latent classes has better joint LL, marginal LL, AIC, and average prediction LL, than both LCCM and GM-LCCM with two latent classes. However, it comes as no surprise that the LCCM with two latent classes has the lowest BIC. This is due to the nature of the GMM and its different covariance structures which result in higher number of parameters for the proposed GM-LCCM.

**Table 7: Explanatory variables used in the models**

| Variable | Type | Description | Sub-Model |
|---|---|---|---|
| $Cost_{ST}$ | Continuous variable | Cost of a one-way trip by shared-ride taxi (in 1,000 L.L.)[2] | |
| $Cost_{SH}$ | Continuous variable | Cost of a one-way trip by shuttle including parking cost at the satellite parking (in 1,000 L.L.) | |
| $Cost_{Car}$ | Continuous variable | Fuel and parking cost of a one-way trip by car (in 1,000 L.L.) | Class-specific choice model |
| $TT_{ST}$ | Continuous variable | Travel time of one-way trip by shared taxi (in hours) | |
| $TT_{SH}$ | Continuous variable | Travel time of one-way trip by shuttle including access time to the satellite parking (in hours) | |
| $TT_{Car}$ | Continuous variable | Travel time of one-way trip by car (in hours) | |
| *Head* | Continuous variable | Shuttle headway (in hours) | |
| *Age* | Continuous variable | Age of the respondent (in years/10) | |
| *Grade* | Continuous variable | A number between 1 and 16 used to specify the job, seniority, and salary of a staff member | Class membership model |
| *Cars /Drivers* | Continuous variable | Ratio of number of cars available over number of licensed drivers per household | |
| *Nb* | Continuous variable | Number of people who are usually present in the car during the trip from home to AUB | |

**Table 8: Summary results of LCCM and GM-LCCM**

| | Covariance Type | Nb of Parameters | Joint LL[a] | LL[b] | AIC | BIC | Pred. LL |
|---|---|---|---|---|---|---|---|
| **LCCM (K=2)** | | 47 | | -4,910.92 | 9,915.84 | **10,191.41** | -1,024.93 |
| **GM-LCCM (K=2)** | Full | 71 | -8,476.35 | -4,937.64 | 10,017.28 | 10,433.57 | -1,018.10 |
| | Tied | 61 | -8,533.22 | -4,911.08 | 9,944.16 | 10,301.82 | -1,012.62 |
| | Diagonal | 59 | -8,564.64 | -4,935.51 | 9,989.02 | 10,334.95 | -1,017.87 |
| | Spherical | 53 | -8,575.90 | -4,927.54 | 9,961.08 | 10,271.83 | -1,016.51 |
| **GM-LCCM (K=3)** | Spherical | 80 | **-7,042.21** | **-4,893.29** | **9,946.58** | 10,415.64 | **-998.41** |

a: joint log-likelihood of the GM-LCCM

b: marginal log-likelihood of the GBMLCCM and LCCM

---

[2] 1 USD = 1,500 Lebanese Lira (L.L.) at the time the survey was conducted.

**Table 9: LCCM – K = 2**

| Variable | Class 1 | Class 2 |
|---|---|---|
| | Class-specific choice model | |
| $C_{Car1}$ | 0.3717 (0.00) | -2.562 (0.00) |
| $C_{Car2}$ | 0.2981 (0.01) | -2.057 (0.00) |
| $C_{Car3}$ | 0.516 (0.00) | -2.356 (0.00) |
| $C_{Car4}$ | -0.422 (0.01) | -3.087 (0.00) |
| $C_{ST1}$ | -0.464 (0.00) | -1.601 (0.03) |
| $C_{ST2}$ | -0.172 (0.24) | -2.099 (0.00) |
| $C_{ST3}$ | -0.108 (0.61) | -1.054 (0.03) |
| $C_{ST4}$ | -0.347 (0.25) | -3.081 (0.00) |
| $C_{ST5}$ | -0.217 (0.53) | -0.158 (0.53) |
| $C_{SH1}$ | -0.280 (0.02) | -2.255 (0.00) |
| $C_{SH2}$ | 0.413 (0.00) | -2.989 (0.00) |
| $C_{SH3}$ | 0.678 (0.00) | -2.259 (0.00) |
| $C_{SH4}$ | 0.373 (0.09) | -3.93 (0.00) |
| $C_{SH5}$ | 0.378 (0.16) | -1.522 (0.00) |
| $\beta_{Cost\_Car}$ | -0.0456 (0.00) | -0.0446 (0.00) |
| $\beta_{Cost\_ST}$ | -0.109 (0.00) | -0.101 (0.00) |
| $\beta_{Cost\_SH}$ | -0.0998 (0.00) | -0.0400 (0.00) |
| $\beta_{Time\_Car}$ | -0.658 (0.00) | -0.409 (0.00) |
| $\beta_{Time\_ST}$ | -0.646 (0.00) | -0.372 (0.00) |
| $\beta_{Time\_SH}$ | -0.387 (0.00) | -0.252 (0.00) |
| $\beta_{Head}$ | -0.565 (0.00) | -0.0423 (0.65) |
| **Variable** | **Class membership model** | |
| $ASC$ | - | -2.271 (0.00) |
| $\beta_{Grade}$ | - | 0.0569 (0.00) |
| $\beta_{C/D}$ | - | 0.267 (0.37) |
| $\beta_{Age}$ | | 0.587 (0.00) |
| $\beta_{Nb}$ | - | 0.0850 (0.26) |
| **VOT ($/hr)** | | |
| $Car$ | 9.61 | 6.11 |
| $ST$ | 3.96 | 2.44 |
| $SH$ | 2.59 | 4.20 |

**Table 10: GM-LCCM – K = 2**

| Variable | Class 1 | Class 2 |
|---|---|---|
| | Class-specific choice model | |
| $C_{Car1}$ | 0.361 (0.00) | -2.502 (0.00) |
| $C_{Car2}$ | 0.290 (0.01) | -2.042 (0.00) |
| $C_{Car3}$ | 0.508 (0.00) | -2.388 (0.00) |
| $C_{Car4}$ | -0.430 (0.00) | -3.077 (0.00) |
| $C_{ST1}$ | -0.465 (0.00) | -1.617 (0.04) |
| $C_{ST2}$ | -0.174 (0.24) | -2.085 (0.00) |
| $C_{ST3}$ | -0.108 (0.61) | -1.075 (0.03) |
| $C_{ST4}$ | -0.347 (0.25) | -3.154 (0.00) |
| $C_{ST5}$ | -0.209 (0.55) | -0.159 (0.53) |
| $C_{SH1}$ | -0.286 (0.02) | -2.295 (0.00) |
| $C_{SH2}$ | 0.403 (0.00) | -3.029 (0.00) |
| $C_{SH3}$ | 0.661 (0.00) | -2.290 (0.00) |
| $C_{SH4}$ | 0.354 (0.11) | -4.016 (0.00) |
| $C_{SH5}$ | 0.379 (0.15) | -1.521 (0.00) |
| $\beta_{Cost\_Car}$ | -0.0462 (0.00) | -0.0442(0.00) |
| $\beta_{Cost\_ST}$ | -0.110 (0.00) | -0.101 (0.00) |
| $\beta_{Cost\_SH}$ | -0.0993 (0.00) | -0.0401 (0.00) |
| $\beta_{Time\_Car}$ | -0.653 (0.00) | -0.409 (0.00) |
| $\beta_{Time\_ST}$ | -0.641 (0.00) | -0.372 (0.00) |
| $\beta_{Time\_SH}$ | -0.384 (0.00) | -0.252 (0.00) |
| $\beta_{Head}$ | -0.561 (0.00) | -0.0442 (0.64) |
| **Variable** | **Class membership model** | |
| $\pi$ | 0.425 | 0.575 |
| $\mu_{Grade}$ | -0.3034 | 0.2246 |
| $\mu_{C/D}$ | -0.062 | 0.0459 |
| $\mu_{Age}$ | -0.4087 | 0.3025 |
| $\mu_{Nb}$ | -0.0693 | 0.0513 |
| **VOT ($/hr)** | | |
| $Car$ | 9.42 | 6.16 |
| $ST$ | 3.90 | 2.45 |
| $SH$ | 2.58 | 4.19 |

### Table 11: GM-LCCM – K = 3

| Variable | Class 1 | Class 2 | Class 3 |
|---|---|---|---|
| | **Class-specific choice model** | | |
| $C_{Car1}$ | -2.238 (0.00) | 0.444 (0.00) | -0.102 (0.78) |
| $C_{Car2}$ | -1.818 (0.00) | 0.290 (0.02) | 0.327 (0.25) |
| $C_{Car3}$ | -2.105 (0.00) | 0.677 (0.00) | -0.0412 (0.88) |
| $C_{Car4}$ | -2.897 (0.00) | -0.224 (0.18) | -1.263 (0.00) |
| $C_{ST1}$ | -1.608 (0.01) | -0.331 (0.01) | -1.105 (0.00) |
| $C_{ST2}$ | -2.305 (0.00) | -0.00930 (0.95) | -0.907 (0.03) |
| $C_{ST3}$ | -1.071 (0.02) | 0.0262 (0.91) | -0.938 (0.09) |
| $C_{ST4}$ | -3.231 (0.00) | -0.073 (0.82) | -1.778 (0.08) |
| $C_{ST5}$ | -0.181 (0.47) | -0.284 (0.48) | -0.244 (0.76) |
| $C_{SH1}$ | -2.192 (0.00) | -0.240 (0.07) | -0.557 (0.08) |
| $C_{SH2}$ | -3.170 (0.00) | 0.556 (0.00) | -0.535 (0.11) |
| $C_{SH3}$ | -2.12 (0.00) | 0.862 (0.00) | -0.959 (0.03) |
| $C_{SH4}$ | -4.382 (0.00) | 0.627 (0.01) | -1.627 (0.00) |
| $C_{SH5}$ | -1.450 (0.00) | 0.601 (0.04) | -1.149 (0.05) |
| $\beta_{Cost\_Car}$ | -0.0451 (0.00) | -0.0707 (0.00) | -0.0172 (0.00) |
| $\beta_{Cost\_ST}$ | -0.0991 (0.00) | -0.107 (0.00) | -0.123 (0.11) |
| $\beta_{Cost\_SH}$ | -0.0421 (0.00) | -0.0832 (0.00) | -0.120 (0.01) |
| $\beta_{Time\_Car}$ | -0.410 (0.00) | -0.717 (0.00) | -0.614 (0.00) |
| $\beta_{Time\_ST}$ | -0.384 (0.00) | -0.777 (0.00) | -0.349 (0.26) |
| $\beta_{Time\_SH}$ | -0.259 (0.00) | -0.519 (0.00) | -0.107 (0.00) |
| $\beta_{Head}$ | -0.0114 (0.91) | -0.757 (0.00) | -0.380 (0.06) |
| **Variable** | **Class membership model** | | |
| $\pi$ | 0.570 | 0.339 | 0.091 |
| $\mu_{Grade}$ | 0.258 | -0.172 | -0.981 |
| $\mu_{C/D}$ | 0.0456 | -0.222 | 0.540 |
| $\mu_{Age}$ | 0.329 | -0.314 | -0.897 |
| $\mu_{Nb}$ | 0.0778 | 0.0861 | -0.810 |
| | ***VOT*** ($/*hr*) | | |
| $Car$ | 6.07 | 6.76 | 23.81 |
| $ST$ | 2.58 | 4.85 | 1.89 |
| $SH$ | 4.10 | 4.16 | 0.60 |

# 5   CONCLUSION

In this paper, we investigated the feasibility of combining Gaussian-Bernoulli Mixture Models with Latent Class Choice Models. We contributed to the literature by formulating and estimating simultaneously a semi-nonparametric demand model that combines unsupervised machine learning (mixture models) and econometric models without jeopardizing the behavioral and economic interpretability of the choice models. The proposed hybrid model is inspired by LCCMs which include two sub-components: a class membership model and a class-specific choice model. The class membership model, which predicts the probability of a decision-maker belonging to a specific latent class/cluster, is formulated as a mixture with Gaussian and Bernoulli distributions instead of a random utility formulation. Conditioned on the class assignments, the class-specific choice model formulates the probability of a particular alternative being chosen by a decision-maker. The model was tested and compared to the traditional MNL, continuous mixed logit, and LCCM using a revealed preferences case study on travel mode choice behavior. The model was also tested and compared to LCCM using a stated preferences case study on weekly frequencies of commuting by different modes. Results showed that the GBM-LCCM is capable of capturing more complex taste heterogeneity than the traditional LCCM by identifying a larger number of latent classes. This might be due to the fact that mixture models allow more flexibility than the linear-in-the-parameters utility specification of the latent classes. In addition, it is capable of improving the prediction accuracy of the choice models. These improvements are accomplished without any interpretability losses, neither at the class membership level nor at the class-specific choice model level. In fact, the latent classes can be easily interpreted and marginal effects in addition to economic indicators (e.g. willingness to pay) can be directly inferred from the model. To sum up, this new approach satisfies the main properties of an effective econometric behavioral model, as set by McFadden.

However, this study is not devoid of limitations. There are several extensions that should be explored in future work. First, the Gaussian-Bernoulli mixture model assumes that the continuous-binary variables that are used for clustering are uncorrelated. Although the Gaussian part of the mixture model offers different covariance structures for the continuous set of variables, future work should explore ways to capture correlations between all continuous-binary variables of the class membership model. A second straightforward extension could be related to within-class heterogeneity. Previous studies have shown that individuals with similar socio-economic characteristics, and thus belonging to the same latent class, might not have the same preferences or taste homogeneity (Bujosa et al., 2020). Therefore, a natural extension of the GBM-LCCM is to integrate random distributions or mixture of random distributions of taste coefficients within the class-specific choice models. Third, although two different types of datasets have been used and several specifications in addition to a logarithmic transformation have been tested, it would be worthwhile to investigate whether the findings of this study generalize to different applications, specifications, and attribute transformations. Finally, though the focus of the two applications was on travel mode choice, the model can be applied to any application with a finite discrete choice set where it is believed that taste heterogeneity exists among decision-makers. It is hoped that these extensions could provide a stronger evidence base for the potential merits of the proposed framework to the choice modeling community and transportation planners.

## Acknowledgments

This research did not receive any specific grant from funding agencies in the public, commercial, or not-for-profit sectors. However, the first author received funding for his research from the Maroun Semaan Faculty of Engineering and Architecture at the American University of Beirut (AUB) and from the Department of Management Engineering at the Technical University of Denmark (DTU). An earlier version of this paper was presented at the 99[th] Transportation Research Board Annual Meeting in Washington, D.C. in January 2020.